\documentclass{ws-procs9x6}

\begin{document}

\title{High $Q^2$ Structure Functions and Parton Distributions}

\author{B. PORTHEAULT
}

\address{Laboratoire de l'Acc\'el\'erateur Lin\'eaire \\
Univertist\'e Paris Sud,\\ 
F-91898 Orsay Cedex\\ 
E-mail: portheau@lal.in2p3.fr}

\maketitle

\abstracts{This contribution reviews the main achievements in inclusive measurements made by the H1 and ZEUS collaborations during the first phase of HERA data taking. The QCD analysis of these data by both collaborations are described. The case for a common QCD analysis is  briefly discussed, with an emphasis on the possible  $W$ mass extraction.}

At leading order in the electroweak (EW) interaction, the double differential cross section of inclusive Deep Inelastic Scattering (DIS) can be expressed in terms of structure functions

\begin{equation}
\frac{\mathrm{d}^2\sigma_{NC}^{\pm}}{\mathrm{d}x\mathrm{d}Q^2}=\frac{2\pi \alpha^2}{xQ^{4}}\left[ Y_{+}\tilde{F}_{2}-y^2\tilde{F}_{L}\mp Y_{-}x\tilde{F}_{3} \right],
\label{NC}
\end{equation}
for Neutral Currents (NC) where $Y_{\pm}=1\pm (1-y)^2$, and similarly for Charged Current (CC)
\begin{equation}
\frac{\mathrm{d}^2\sigma_{CC}^{\pm}}{\mathrm{d}x\mathrm{d}Q^2}=\frac{G_{F}^{2}}{4\pi x}\left[\frac{M_{W}^{2}}{Q^2+M_{W}^{2}} \right]^2 \left[ Y_{+} F_{2}^{CC\pm}-y^{2}F_{L}^{CC\pm}\mp Y_{-}x F_{3}^{CC\pm} \right],
\label{CC}
\end{equation}
where the structure functions exhibit a dependency upon the incoming lepton charge. The QCD factorisation theorem allows the separation of the long distance physics and the short distance physics, such that the structure functions can be expressed as convolutions of universal parton distributions (pdfs) and perturbatively computable kernels. 

Rich physics can be extracted with the measurement of high $Q^2$ inclusive cross sections. On one hand the short distance physics can be tested. For example, the data can be used to test  the structure of the EW interaction, and to search  for new physics. On the other hand, the long distance physics can be measured. Through suitable combinations of cross section or with the help of a prediction, the structure functions can be extracted out of Eqs. (\ref{NC}) and (\ref{CC}). The case of the $F_{L}$ structure function is discussed in \cite{vladimirFL}. The QCD analyses (the so-called ``QCD fits'') aim at extracting  the pdfs through the  QCD evolution.
It is also possible to  extract  any parameter entering in the expression of the cross section. The best example of such an extraction is $\alpha_{s}$, but in principle it also works for EW parameters.

In this contributions, inclusive cross section measurements and structure function extractions are reviewed.  The QCD analysis done by the H1 and ZEUS collaborations are then detailed. Finally the case for a common QCD analysis is presented together with a strategy for an extraction of the $W$ mass.

\section{Inclusive measurement results}

\begin{figure}[htbp]
\centerline{\epsfxsize=9cm\epsfbox{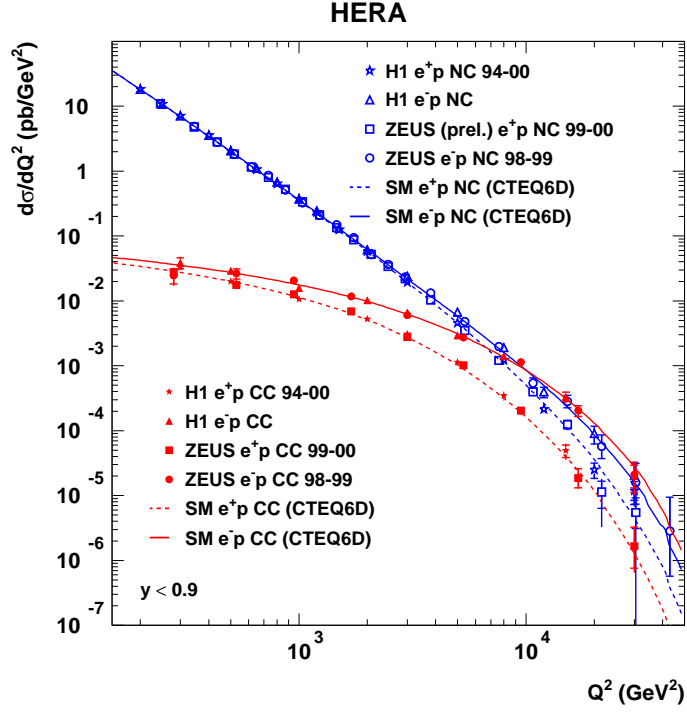}}   
\caption{Neutral and Charged Current cross sections measured by the H1 and ZEUS collaborations. \label{fig:ewplot}}
\end{figure}

The inclusive  differential cross sections as a function of $Q^2$ is shown in Fig. \ref{fig:ewplot} (see\cite{H1NcCcQCDfit,H1NcCcEm,H1NcCcEp,ZeusCcEp,ZeusNcEp,ZeusCcEm,ZeusNcEm,ZeusCcEp00,Zeus9900NC}). 
Important information can be extracted from this figure. At low-medium  $Q^2$, where the NC cross section is the largest, the precision on the NC cross section is very high. At very large $Q^2$ where the $\gamma-Z^{0}$ interference and $Z^{0}$ exchange start to play a non-negligible role, the $e^{+}$ and $e^{-}$ cross sections start clearly to differ due to the opposite sign of the $xF_{3}$ contribution, which then can be extracted. For the CC cross section, the difference between the  $e^{+}$ and $e^{-}$ cross sections is clear, reflecting  differences in the parton distributions probed and the different helicity factors. At $Q^{2}\simeq M_{W}^{2}$ the CC and NC cross sections become of the same magnitude, which illustrates the deep relationship between the normalisation constants: $G_{F}^{2}/16\pi \sim 2\pi\alpha^{2}/M_{W}^{4}$. This is the HERA manifestation of EW unification. 

\subsection{Neutral Current}

\begin{figure}[htbp]
\centerline{\epsfxsize=10cm\epsfbox{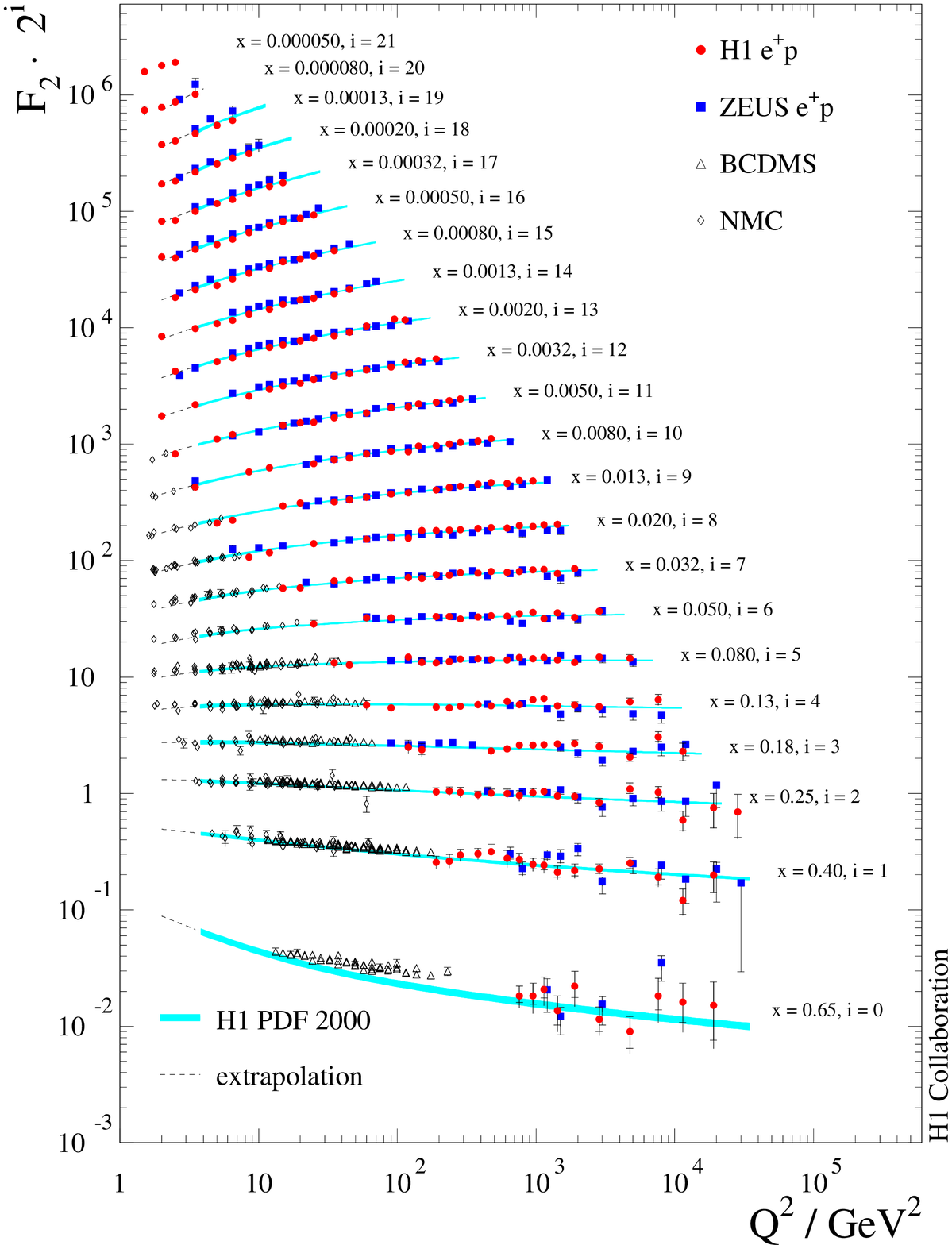}}   
\caption{$F_{2}$ structure function measured by the H1 and ZEUS collaboration together with the H1PDF2000 fit. Note that the fixed target data shown are not included in the fit. \label{fig:qcdplot}}
\end{figure}

The $F_{2}$ structure function is shown in  Fig. \ref{fig:qcdplot}. It exhibits the QCD pattern of the scaling violations over several orders of magnitude in $x$ and $Q^2$.
The typical accuracy is 2--3\% in most of the phase space. At high $x$ and $Q^2$, the potentially interesting region to look for exotics, one sees the statistically limited precision of the measurement compared to the fixed target experiments. 



\begin{figure}[htbp]
\centerline{\epsfxsize=6cm \epsfysize=5cm \epsfbox{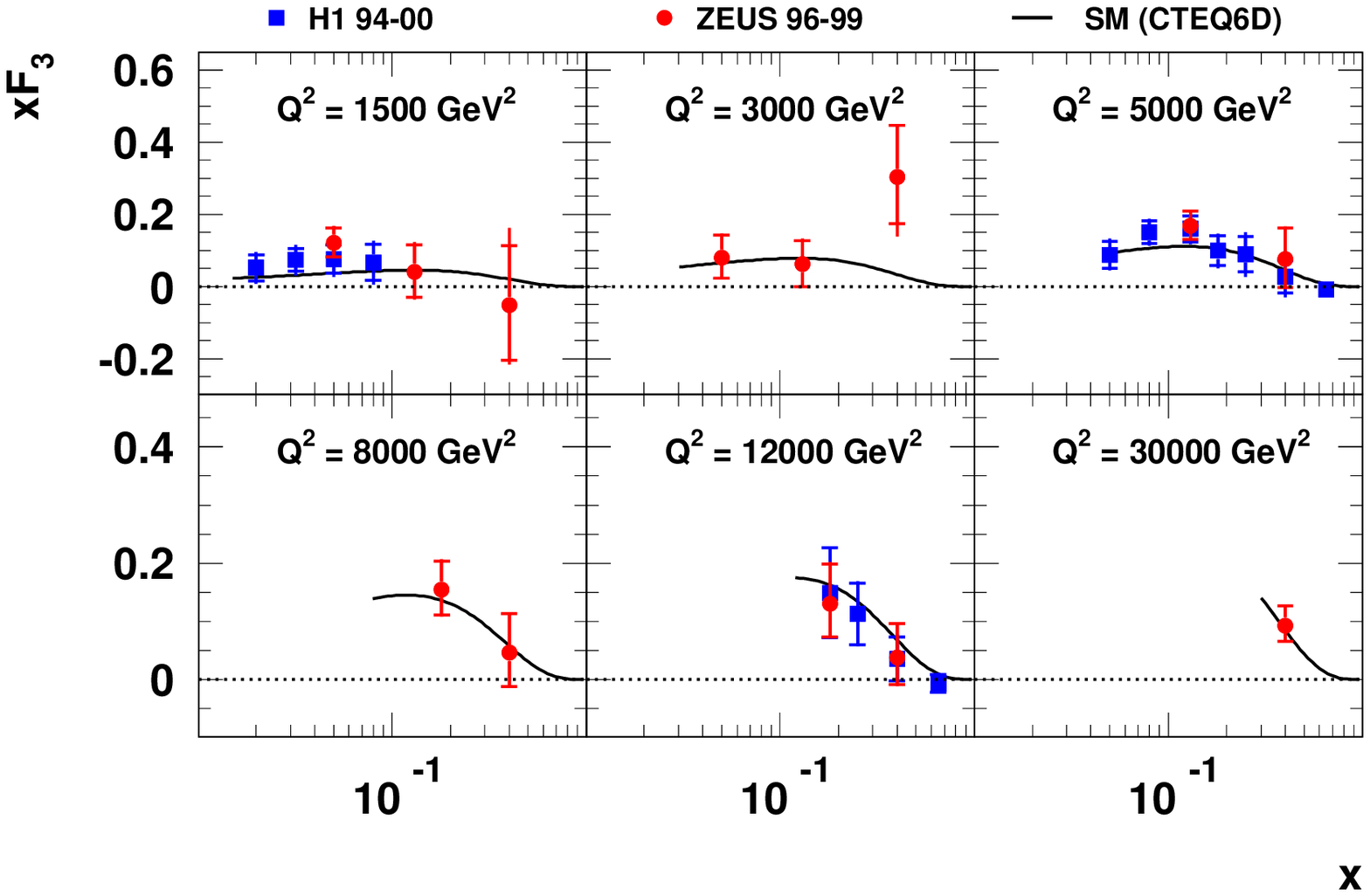}\epsfxsize=6cm \epsfysize=5cm\epsfbox{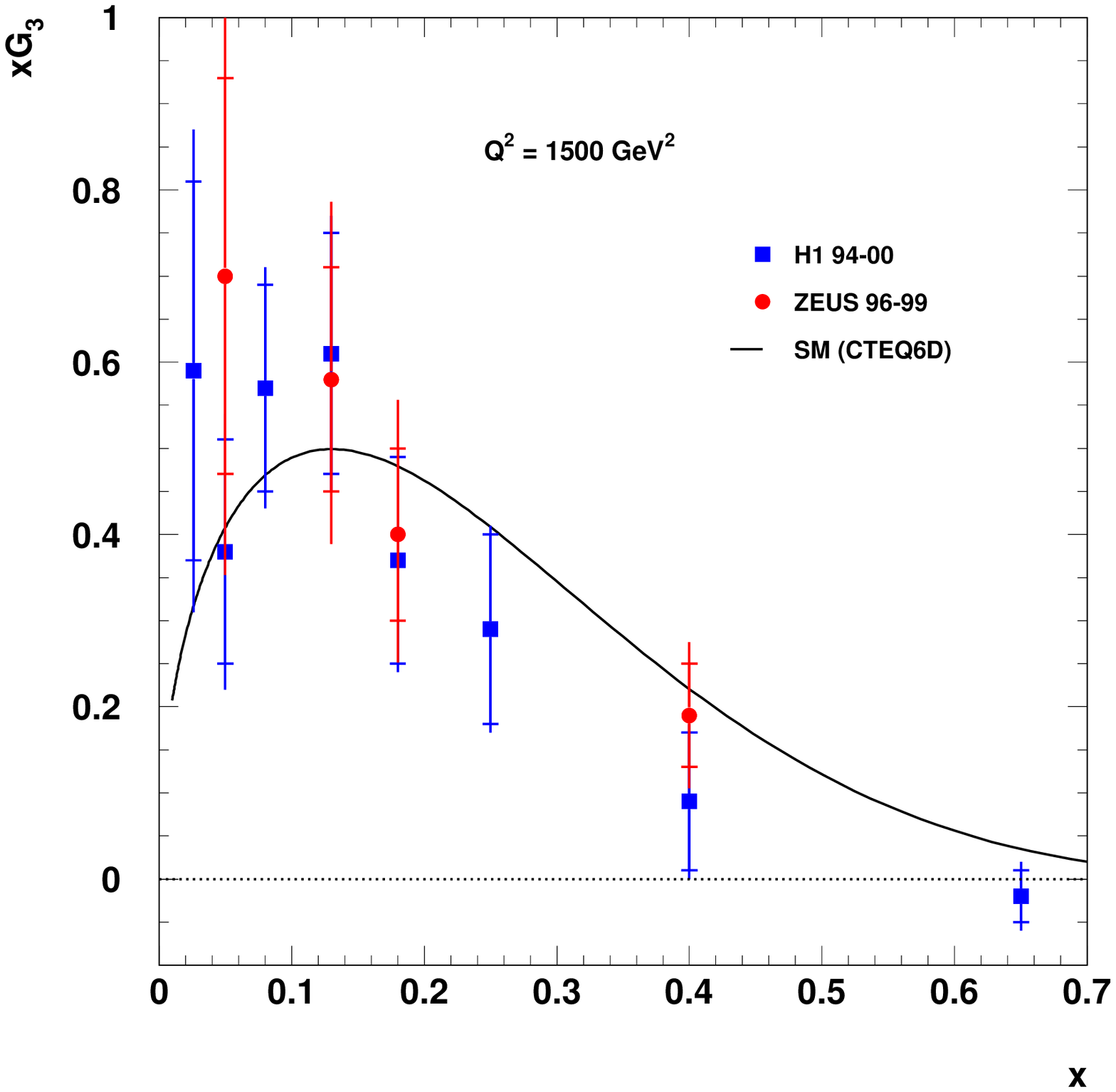}}    
\caption{Left: $x F_{3}$ structure function. Right: $x F_{3}^{\gamma Z}$ structure function.  \label{fig:xf3}}
\end{figure}

The visible differences in the $e^{+}$ and $e^{-}$ cross sections at high $x$ allows the extraction of the $xF_{3}$ structure function. The results are shown in Fig. \ref{fig:xf3} in the left plot. 
As $xF_{3}=-a_{e}\chi_{Z} x F_{3}^{\gamma Z}+2a_{e}v_{e}\chi_{Z}^{2}x F_{3}^Z$ where $\chi_{Z}\propto Q^{2}/(Q^2+M_{Z}^2)$, the propagator contribution and the small value of $v_{e}$ have the effect that the $x F_{3}^Z$ contribution to $xF_{3}$ is always below 3\% in the measured range. This allows the extraction of $xF_{3}^{\gamma Z}$  as shown in the right plot of Fig \ref{fig:xf3}. As $xF_{3}^{\gamma Z}\propto 2x u_{val}+x d_{val}$ at leading order, this provides an access to the valences parton distributions. However there is still a large  error of about 30\% due mainly to the limited statistics of the $e^{-}$ data sample available.

\subsection{Charged Current}

\begin{figure}[htbp]
\centerline{\epsfxsize=10cm\epsfbox{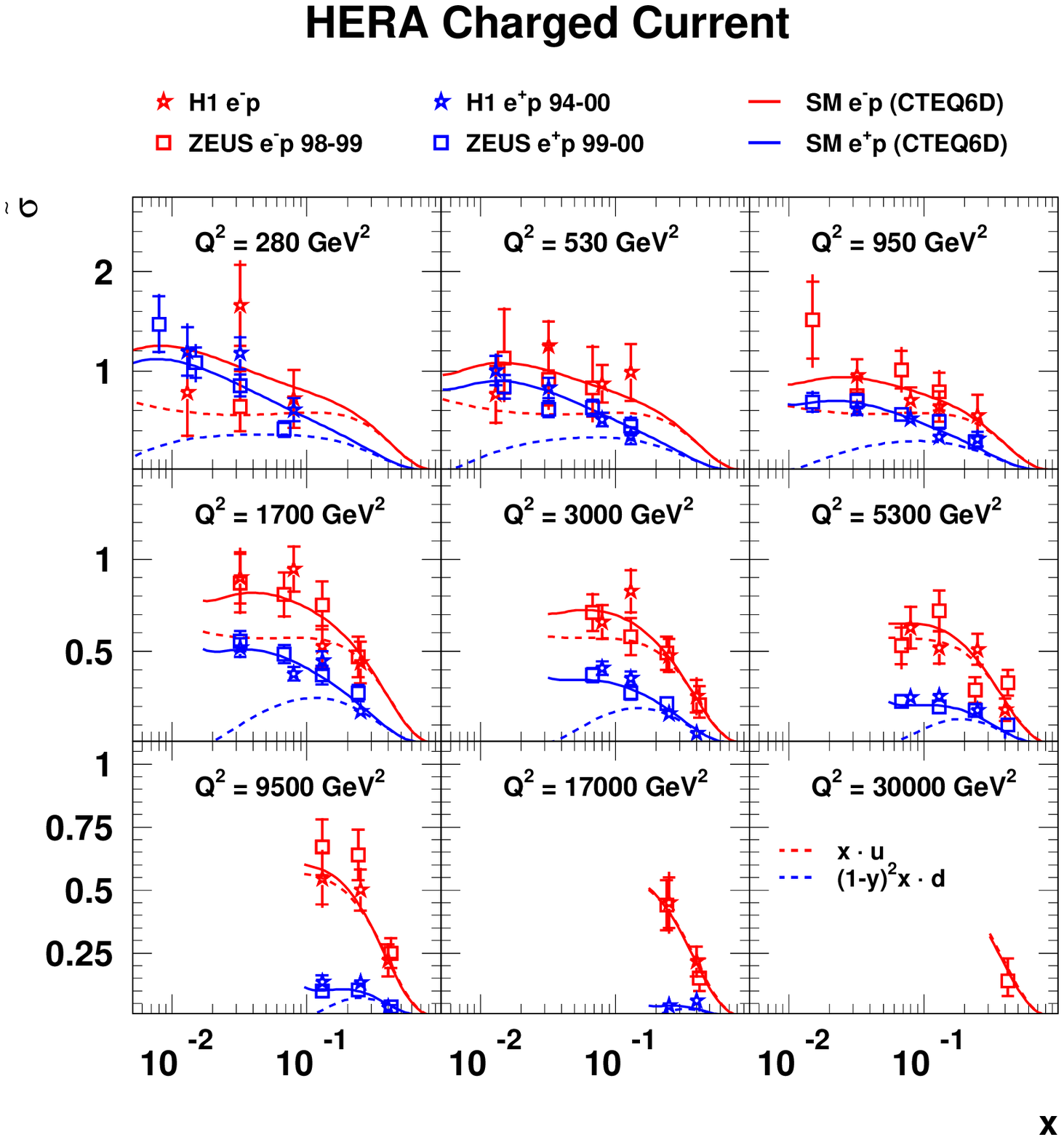}}   
\caption{Charged Current reduced cross section for $e^+$ and $e^{-}$ data. \label{fig:CC}}
\end{figure}

For the Charged Current process one can  define a reduced cross section
\begin{equation}
\tilde{\sigma}_{CC}^{\pm}=\frac{2\pi x}{G_{F}^{2}}\frac{(Q^{2}+M_{W}^{2})^2}{M_{W}^{4}}\frac{\mathrm{d}^2\sigma_{CC}^{\pm}}{\mathrm{d}x\mathrm{d}Q^2}.
\end{equation}
The results of the H1 and ZEUS collaborations are shown on Fig. (\ref{fig:CC}) for the $e^+$ and $e^{-}$ data sets.

\begin{figure}[htbp]
\centerline{\epsfxsize=9cm\epsfbox{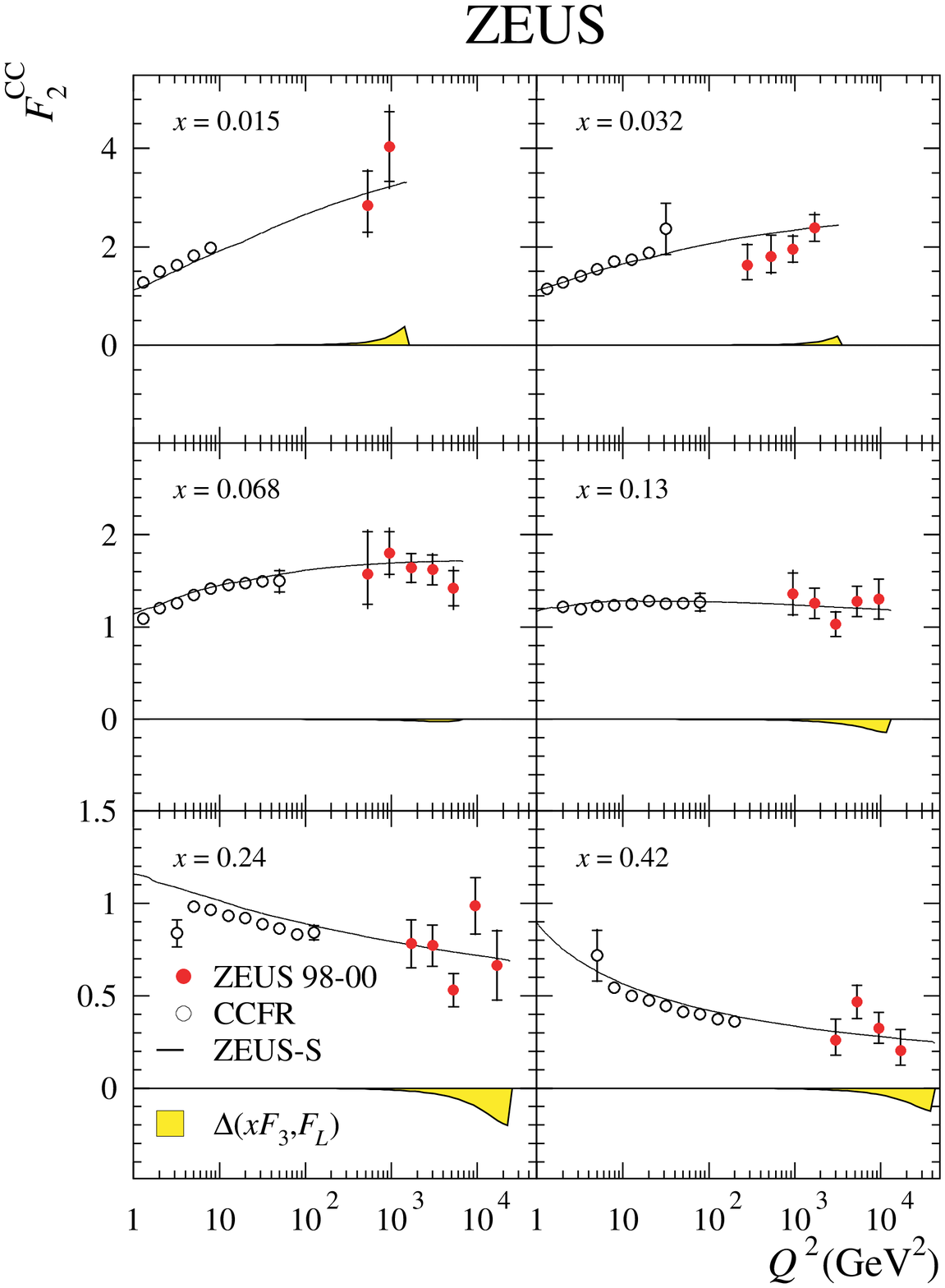}}   
\caption{$F_{2}^{CC}$ extraction with the ZEUS CC data and the ZEUS--S fit. \label{fig:F2CC}}
\end{figure}

The good understanding of the hadronic response of the detectors has led to  typical systematic errors of the order of 6\% for the 99--00 data set. The total error is clearly dominated by statistics at the largest $x$ and $Q^2$. At leading order the reduced cross sections reads $\tilde{\sigma}_{CC}^{+}=x\left[ \bar{u}+\bar{c}+(1-y)^2(d+s) \right]$ and  $\tilde{\sigma}_{CC}^{-}=x\left[ u+c+(1-y)^2(\bar{d}+\bar{s}) \right]$. This decomposition is useful to see the contribution of u--type quarks and d--type quarks to the total reduced cross section. One can see in Fig. \ref{fig:CC} that at large $x$ the $e^+$ cross section is dominated by the $d$ quark distribution, and the $e^{-}$ cross section is dominated by the $u$ quark distribution. These data provide an important constraints at high $x$ for the QCD analyses and are necessary for flavour separation of parton distributions.
 Combining the $e^{+}$ and $e^{-}$ data, the ZEUS collaboration performed an extraction  of the structure function $F_{2}^{CC}$ defined by
$F_{2}^{CC}=F_{2}^{CC+}+F_{2}^{CC-}$. It is obtained with
\begin{equation}
F_{2}^{CC}=\frac{2}{Y_{+}}(\tilde{\sigma}_{CC}^{+}+\tilde{\sigma}_{CC}^{-})+\Delta(x F_{3}^{CC\pm}, F_{L}^{CC\pm}),
\end{equation}
where the correction $\Delta(x F_{3}^{CC\pm}, F_{L}^{CC\pm})$ is obtained with the ZEUS--S fit.
 This reads $F_{2}^{CC}\propto u+d+s+\bar{u}+\bar{d}+\bar{s}$ which is similar to the NC expression but with an equal weight for each parton density. The result is shown in Fig. \ref{fig:F2CC}.
This result extends by two orders of magnitude the results of CCFR\cite{CCFRF2cc}.

\section{QCD analysis and extraction of parton distributions}

The precise  data measured allow the fit of flavour separated parton distributions. All the fit details can be found in \cite{H1NcCcQCDfit} and \cite{ZeusQCDfit}. Rather than describing these details here,  the main ideas and data sets used in both fits will be briefly reviewed. The emphasis will be placed only on these HERA parton distributions, and a few relevant technical details will be underlined. The possibility of a combined fit is briefly discussed.

 \subsection{The H1 and ZEUS QCD Fits}
The ZEUS collaboration performed two fits. The first is the ZEUS--S fit, which uses the ZEUS 96--97 $e^{+}$ NC data with BCDMS\cite{BCDMSF2p}, NMC\cite{NMCF2pAndF2d}, E665 proton $F_2$ data\cite{E665F2pAndF2d}. Deuterium data from E665 and NMC are also used with $F_{2}^{D}/F_{2}^{p}$ results  from NMC\cite{NMCF2dSurF2p} and  CCFR $xF_3$ iron data\cite{CCFR}. These data sets provide all the necessary constraints to extract flavor separated parton distributions. The ZEUS--O fit focus on the ZEUS data and uses the NC and CC ZEUS $e^{+}$ and $e^{-}$ data up to the 99 $e^{-}$ data set. This is also the approach used in the H1PDF2000 fit which uses all the H1 HERA--I  NC and CC data in addition to low $Q^2$ data from 96--97 \cite{H1lowq9697}. For the use of only HERA data one has to solve technical problems due to the limitation of constraints. For the ZEUS--O fit this is achieved by arbitrarily fixing parameters to the values of the ZEUS--S fit. For the H1PDF2000 fit a novel ansatz of decomposition has been adopted, and only internal constraints between parameters have been used. Note that in any case assumptions have to be made which are the price to pay for the  pdfs extraction using only HERA data.

Both fits handle  the correlation of  systematic errors in their parameters and errors estimations. Whereas ZEUS use the so-called offset method\cite{zeusErr}, H1 use the Pascaud-Zomer method\cite{h1Err}. The gist is that only a proper treatment of the correlated systematic errors allows the application of the ``standard'' $\Delta \chi^{2}=1$ statistical criteria for error estimate\cite{errorsFit}. This is a major advantage with respect to the global analysis where the use of data sets providing only a total  systematic error for each data point spoils the use of standard statistical tools. So the use of  HERA data alone  made  possible  a precise extraction with reliable error determination in the QCD fit.

The parametrisation of the input pdf at $Q_{0}^{2}$ is also a very delicate issue. The form generally adopted is $xf(x,Q_{0}^{2})=Ax^{B}(1-x)^C P(x)$, where $P(x)$ may take several polynomial-like forms. It is not  trivial to find  a functional form for $P(x)$ such that the fit is flexible enough to ensure a good $\chi^2$, whilst avoiding instabilities due to over--parametrisation. In the end, one has to keep in mind that any choice of parametrisation is more or less arbitrary, and that the distributions and their error depend on the parametric form chosen.

\begin{figure}[htbp] 
\centerline{\epsfxsize=8cm\epsfbox{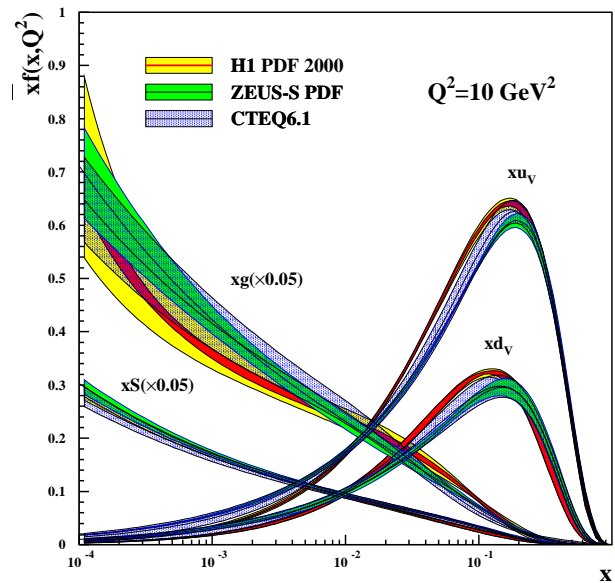}}   
\caption{Parton distributions of the H1PDF2000 and ZEUS--S fits at 10 GeV$^2$ and comparison with the CTEQ6 fit.\label{fig:pdfs}}
\end{figure}

The results are shown in Fig. \ref{fig:pdfs}. The agreement is reasonable between the different fits given the different data sets and the different fitting schemes. From the H1PDF2000 the $u$--type and $d$--type quark distributions precisions are respectively 1\% and 2\% for $x=10^{-3}$,  7\% and 30\% for $x=0.65$. From the ZEUS--S fit the sea distribution precision is 5\% between $x=10^{-4}$ and $10^{-1}$.  

The extraction of parton densities from HERA data alone is a major achievement but the fits are at the limit of the technical fitting possibilities. This is why the use of all the combined H1 and ZEUS data could help to gain in flexibility, in particular to relax some of the assumptions made in the fitting ansatz. However one has to keep in mind that  the combination would help, it may also trigger other problems as the elaboration of a fit is very delicate.

\subsection{Possibilities  of a $W$ mass extraction using a HERA fit.}

\begin{figure}[htbp] 
\centerline{\epsfxsize=6.5cm\epsfbox{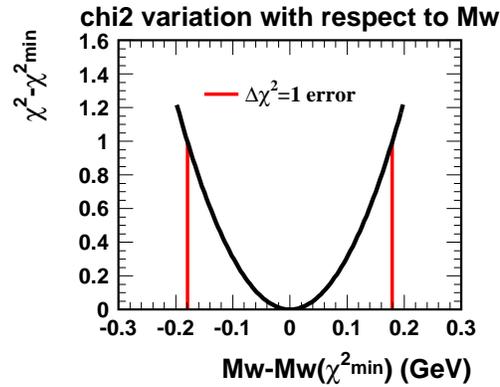}}   
\caption{$\chi^2$ shape obtained with a combined H1 and ZEUS QCD fit for variations of the $W$ mass. \label{fig:mw}}
\end{figure}

To confront the Standard Model with experimental data one needs to specify several parameters that enter in physical quantities. Several schemes are possible: the On Mass Shell (OMS) scheme uses masses as an input ($\alpha,M_{W},M_{Z},M_{H}$) whereas the Modified OMS scheme uses the Fermi constant $G_{F}$ instead of the $W$ mass. The two schemes are related by the relation  
\begin{equation}
G_{F}=\frac{\pi \alpha}{\sqrt{2}\left( 1 - \frac{M_{W}^{2}}{M_{Z}^{2}} \right)}\times \frac{1}{1-\Delta r (\alpha,M_{W},M_{Z},M_{H},m_{top})}
\end{equation}
where the radiative correction $\Delta r$ is a function of the other parameters. The normalisation of the CC cross section depends on the scheme (besides the couplings of leptons to the $Z^{0}$ and its propagator normalisation). So several fitting strategies are possible. It is possible to fit  $M_{W}$ to the CC cross section as a ``propagator mass'', and this has been used several times by H1 and ZEUS, or it is possible to fit  $M_{W}$ in the OMS scheme as a propagator  mass which also fixes the normalisation, assuming the SM is valid. This strategy was proposed in \cite{SpiesbergerRingberg}. A breakthrough to reduce the uncertainty due to the proton structure is the use of a combined QCD--EW fit. This possibility is investigated by using the H1 and ZEUS data (up to the 99--00 ZEUS data set) in a QCD fit using the fitting scheme of the H1PDF2000 fit. The principle is the following: a scan of $W$ mass values is realized, and a full QCD fit is done for each $W$ mass. The total $\chi^{2}$ relative to its minimal value $\chi^{2}_{min}$ as a function of $M_{W}$ is shown in Fig. \ref{fig:mw}. A one sigma experimental (statistical and systematic)  error  with a $\Delta \chi^{2}=1$ error criterion is used, which is possible due to the careful treatment of the correlated systematic errors.
The error obtained is 190 MeV, which corresponds to  $\delta M_{W} / M_{W}=0.2$\%. The use of $M_{W}$ as entering in the normalisation of the CC cross section allows this significant improvement with respect to previous ``propagator mass'' fits with a few percent precision. However note that the validity of this result is associated to the validity of the QCD analysis itself. In particular some theoretical uncertainty contribution to the total error may be present. 

\section{Summary and outlook}

Many cornerstone results in DIS have been achieved in the HERA--I phase of data taking. The inclusive NC and CC cross sections have been measured with a good accuracy, although there is scope for considerable improvements using higher statistics and longitudinal lepton polarisation. These data have already been used in many QCD fits. However there is still an unexploited potential in a combined QCD analysis of the H1 and ZEUS data. This combination could settle many important physics issues such as the $\alpha_{s}$ value, or the determination of the gluon distribution. There is still the possibility of QCD--EW combined fits that could be used from now on to deliver the  HERA physics message on parton distributions and Standard Model parameters.

\section*{Acknowledgments}

The author wishes to thank his colleagues in the H1 and ZEUS collaborations for the measurements presented in this paper. He also thanks the organisers of the workshop for the invitation and for the unique hospitality at the Ringberg castle.

\end{document}